\documentclass[reprint,
superscriptaddress,
amsmath,
amssymb,
aps,
twocols
]{revtex4-1}

\usepackage{subcaption}
\usepackage{xcolor}
\usepackage{physics}
\usepackage{amssymb}
\usepackage{easyReview}
\usepackage{hyperref}
\usepackage{float}
\usepackage{graphicx}

\begin{document}


\title{Molecular dynamics characterization of the free and encapsidated RNA2\\ of CCMV with the oxRNA model}

\author{Giovanni Mattiotti}
\affiliation{Laboratoire Biologie Functionnelle et Adaptative, CNRS UMR 8251, Inserm ERL U1133, Université Paris Cité, 35 rue Hélène Brion, Paris, France}

\author{Manuel Micheloni}
\affiliation{Department of Physics, University of Trento, via Sommarive, 14 I-38123 Trento, Italy}
\affiliation{INFN-TIFPA, Trento Institute for Fundamental Physics and Applications, Trento, Italy}

\author{Lorenzo Petrolli}
\affiliation{Department of Physics, University of Trento, via Sommarive, 14 I-38123 Trento, Italy}
\affiliation{INFN-TIFPA, Trento Institute for Fundamental Physics and Applications, Trento, Italy}

\author{Samuela Pasquali}
\affiliation{Laboratoire Biologie Functionnelle et Adaptative, CNRS UMR 8251, Inserm ERL U1133, Université Paris Cité, 35 rue Hélène Brion, Paris, France}

\author{Luca Tubiana}
\affiliation{Department of Physics, University of Trento, via Sommarive, 14 I-38123 Trento, Italy}
\affiliation{INFN-TIFPA, Trento Institute for Fundamental Physics and Applications, Trento, Italy}

\author{Raffaello Potestio}
\affiliation{Department of Physics, University of Trento, via Sommarive, 14 I-38123 Trento, Italy}
\affiliation{INFN-TIFPA, Trento Institute for Fundamental Physics and Applications, Trento, Italy}




\begin{abstract}
The cowpea chlorotic mottle virus (CCMV) has emerged as an exemplary model system to assess the balance between electrostatic and topological features of ssRNA viruses, specifically in the context of the viral self-assembly process. Yet, in spite of its biophysical significance, little structural data of the RNA content of the CCMV virion is currently available.
Here, the conformational dynamics of the RNA2 fragment of CCMV was assessed \textit{via} coarse-grained molecular dynamics simulations, employing the oxRNA2 model. The behavior of RNA2 has been characterized both as a freely-folding molecule and within a mean-field depiction of a CCMV-like capsid. For the latter, a multi-scale approach was employed, to derive a radial potential profile of the viral cavity, from atomistic structures of the CCMV capsid in solution. The conformational ensembles of the encapsidated RNA2 were significantly altered with respect to the freely-folding counterparts, as shown by the emergence of long-range motifs and pseudoknots in the former case. Finally, the role of the N-terminal tails of the CCMV subunits (and ionic shells thereof) is highlighted as a critical feature in the construction of a proper electrostatic model of the CCMV capsid.
\end{abstract}

\maketitle


\section{Introduction}

Viruses are nano-sized pathogenic ``organisms'', capable of self-replicating by hijacking the molecular machinery of a host cell -- this process affecting humans, livestock, plants, and bacteria alike~\cite{Holmes2009}. Yet, they constitute a potential technology for pharmaceutical applications, e.g. as vectors for vaccines~\cite{zhang2015biomolecular}: In all respects, viruses are natural nano-machines whose activity has been fine-tuned by evolution, thereby representing a model system to benchmark fundamental biophysical theories on RNA-protein interactions, protein-protein interactions, and on the interplay between evolution and the physico-chemical constraints enforced on simple genomes~\cite{zandi2020virus, bruinsma2021physics}. 

The simplest viruses -- such as the brome mosaic virus (BMV), the cowpea chlorotic mottle virus (CCMV), and the MS2 bacteriophage -- are composed of a handful of proteins describing an icosahedral shell, or capsid, about a long ssRNA molecule.
It has been shown that solutions of viral RNAs and capsid proteins might spontaneously self-assemble into mature, infective viruses \textit{in vitro}, at proper pH and salt concentration~\cite{bruinsma2021physics, fraenkel1955reconstitution, bancroft1970self, adolph1976assembly}. Furthermore, when different RNA molecules are involved, e.g., in a competitive scenario, most viruses are capable to selectively package their homologous RNA in an extraordinarily efficient manner~\cite{comas2012vitro, bruinsma2021physics, zandi2020virus}. 

The self-assembly of viruses has been the object of intense experimental, theoretical, and computational research over the past fifteen years~\cite{zandi2020virus, bruinsma2021physics, hagan2016recent, comas2012vitro, cadena2012self, erdemci2014rna, erdemci2016effects, erdemci2017rna, comas2019packaging}. While diverse aspects are still open for debate, a general picture has emerged, whereby the viral assembly follows a nucleation-and-growth process, mostly driven by non-specific electrostatic forces: Yet, there are evidences \cite{lomonosoff2019} that this process is sometime aided by selective, short-range interactions taking place between highly-conserved regions of RNA, or packaging signals (PSs), and the capsid proteins (CPs). These interactions are critical in both starting the self-assembly and encapsidating the correct (homologous) RNA. For example, the co-assembly between capsid subunits and the unique PS in the RNA of the helical Tobacco Mosaic Virus (TMV) yields a nucleation seed~\cite{saunders2022tobacco}, which further grows into a mature capsid most likely \textit{via} non-specific electrostatic interactions. A more complex scenario unfolds in MS2, with several PSs attaching onto specific sites of the capsid proteins~\cite{Dykeman2013,Dykeman2014,rolfsson2016direct}; these are likely related to the precise organization of RNA in the capsid, which allowed the cryo-EM imaging of more than 90\% of the viral RNA content of MS2~\cite{koning2016MS2}. In other viruses, the existence of PSs has been only hinted (e.g., in BMV), or none have been found altogether -- which is the case of CCMV -- thereby suggesting that the self-assembly process is dominated by non-specific, electrostatic forces between RNA and the CPs~\cite{zandi2020virus, bruinsma2021physics}.

As shown by experimental and theoretical works, specifically on CCMV, viruses are nevertheless expected to discriminate between RNA substrates upon their shape and diameter, even in the absence of selective interactions~\cite{comas2012vitro,cadena2012self}: this capability is accounted for by both the degree of branching of the RNA molecules~\cite{erdemci2014rna,erdemci2016effects,erdemci2017rna} and the interactions between RNA and a set of disordered, positively-charged arginine-rich motifs (ARMs), lining the inner walls of diverse viral capsids~\cite{perlmutter2013viral,dong2020effect}. In fact, it appears that viral RNAs have been subject to evolutionary pressures affecting not only their sequence and conserved folded motifs~\cite{olsthoorn1996,klovins1998,Sanjuan2011,Dykeman2013,Dykeman2014}, but their size, shape, and branching degree alike~\cite{yoffe2008predicting, gopal2014viral, tubiana2015synonymous}. 

CCMV is composed of an icosahedrally-symmetric capsid of 180 identical CPs, showing a radius of about 28 nm and a triangulation number T = 3, according to the Caspar-Klug classification~\cite{Caspar1962}. Its genome is tripartite, i.e., a native CCMV capsid either packages a single copy of RNA1 (about 3000 nt long), RNA2 (about 2800nt), or co-packages a copy of RNA3 (2100nt) and RNA4 (700nt)~\cite{beren2017effect}. Notably, CCMV spontaneously assembles about a variety of polyelectrolytes, such as polystyrene sulfonate, poly-uracyl chains, and other heterologous viral RNAs, making it a model virus to characterize the role of non-specific interactions in the assembly process. In fact, this broad adaptability has driven experimentalists to perform competition assays, whereby different RNA moieties compete for a limited amount of CCMV CPs~\cite{comas2012vitro, beren2017effect}. Somewhat surprisingly, it was shown that the RNAs of CCMV are outcompeted by heterologous RNA molecules of similar length belonging to, e.g., BMV~\cite{comas2012vitro}. Moreover, the best cargo of the CCMV vessel appears to be a fully linear polyelectrolyte, although the latter yielding smaller capsids, compatible with a triangulation number $T = 2$~\cite{beren2017effect}.
Lastly, as with many other viruses, CCMV is  overcharged~\cite{garmann2014role}, i.e., its RNA bears a significantly higher electrical charge than that of its capsid~\cite{belyi2006electrostatic}. This fact has been explained in the light of the interaction between the highly positively charged ARMs and RNA~\cite{perlmutter2013viral}.  

Despite their biophysical significance, the only data on the properties of encapsidated CCMV RNA at our disposal come from either theoretical assessments based on mean-field descriptions of the cargo~\cite{erdemci2014rna, erdemci2016effects, erdemci2017rna} or from coarse-grained (CG) simulations, where the RNA is mapped to (branched) chains of beads~\cite{perlmutter2013viral}. In other viruses, such as MS2 and STMV, the exact position of a large portion of the genome is known from experiments~\cite{larson2001satellite,zeng2012model,koning2016MS2}, thereby enabling for refined and rigorous simulation approaches~\cite{freddolino2006molecular,poblete2021structural};  this protocol, however, has not been applicable to CCMV, where no structural, experimental data on RNA is available.

Yet, with the recent development of coarse-grained models of RNA, such as oxRNA, and the computational speed-up brought by GPU computing, one might now assess the dynamical evolution and the (structural, topological) ensemble properties of viral RNAs at an adequate level of resolution~\cite{oxrna2014, rovigatti2015comparison, matek2015coarse}, as an alternative to conventional approaches of secondary structure prediction, such as RNAfold of the ViennaRNA package~\cite{lorenz2011viennarna}.

In this work, we embrace this approach and employ coarse-grained simulations to explore the conformational dynamics of the RNA2 fragment of CCMV as a freely-folding molecule in solution and within a capsid-like enclosing sphere. For the latter case, we deploy two mean-field depictions of the viral capsid: A first approach relies on an analytical solution of the Poisson-Boltzmann equation for an empty icosahedral shell, whereas in a second approach we use Gauss' theorem to map the charge distribution of the atomistic structure of the CCMV capsid (and the ionic shells thereof) into a spherically-symmetric, radial potential profile. By these means, we assess a decisive role of the electrostatic forces -- from both the saline medium and the capsid -- that reflect in a diverse conformational dynamics and stability of the RNA2 molecules. In spite of a mild confinement process, the encapsidation enforces a significant bias on the structural ensemble of RNA2, enabling stable long-range interactions (such as pseudo-knots) that are not observed in a freely-folding scenario. Lastly, we discuss the capabilities and limitations of adopting effective and mean-field approaches, as well as of the numerical tools at our disposal.

\section{Materials and Methods}

\subsection{The oxRNA2 force field}

oxRNA2 is a coarse-grained force field, firstly developed for applications to RNA nanotechnology, and derived \textit{via} a top-down approach based on the structural, mechanical, and thermodynamic properties of RNA~\cite{vsulc2012sequence, matek2015coarse}. Each nucleotide is associated with a single rigid body and two interaction sites, which take into account hydrogen-bonding and stacking forces between CG sites, the backbone connectivity, and excluded volume interactions. 

The oxRNA2 force field effectively models the major and minor grooving of the RNA helices, as well as a salt-dependent modulation of the electrostatic interactions \textit{via} a Debye-H{\"u}ckel formulation, thereby adapting the dynamics of RNA in implicit solvent to diverse concentrations of monovalent salt \cite{matek2015coarse}. Moreover, it offers a sequence-dependent description of the strength of stacking and (Watson-Crick, wobble) hydrogen-bonding interactions, fitted upon the melting temperature of RNA duplexes and based on the nearest-neighbour models of SantaLucia and Turner~\cite{santalucia1998unified, mathews1999expanded}.

Lastly, despite lacking a proper definition of non-canonical interactions -- such as Hoogsteen and sugar-edge hydrogen bonds, and ribose zippers -- oxRNA2 consistently displays a variety of tertiary motifs, namely the coaxial stacking of RNA helices, kissing-loops, and pseudo-knots \cite{oxrna2014}.

\subsection{MD simulation protocol}
As mentioned in the Introduction section, CCMV is a multi-partite ssRNA virus, i.e., its genome is packed within three separate capsid vessels, all of which are required by the infective process: Out of four CCMV viral fragments, we chose RNA2, whose gyration radius in solution has been estimated by scattering techniques~\cite{marichal2021relationships}; the $2774$-nt sequence of this fragment is detailed in \textbf{Section I} of the Supplementary Material. All MD simulations of the RNA2 molecule were performed employing the oxRNA2 force field, at diverse concentrations of monovalent salt -- namely $0.15$ M and $0.5$ M -- within the single-GPU, native implementation of the oxDNA simulation code~\cite{poppleton2023oxdna, rovigatti2015comparison}.

The adopted MD protocol is subdivided into three stages: 1) a first equilibration and relaxation of the RNA2 molecule from a linear conformation; 2) a confinement stage, which we here dub as \textit{squeezing}, whereby we steadily enclose the RNA2 molecule within a spherical volume that is compliant with the size of the inner cavity of the CCMV capsid; 3) the dynamics of the encapsidated RNA2 within a mean-field depiction of the electrostatic field of the CCMV capsid and the ionic shells thereof. Hence, \textit{ad hoc} simulation protocols were designed as follows, according to the specific requirements of each stage.

\subsubsection*{Structural equilibration of the RNA2 molecule} \label{section:RNA2Relax_protocol}
The CG coordinates of a linear conformation of the RNA2 molecule were obtained by the TacoxDNA software \cite{suma2019tacoxdna}; thus, a twofold equilibration procedure was applied. 

Firstly, we performed a swift relaxation in the NVT ensemble \textit{via} the Bussi thermostat~\cite{bussi2007canonical}, thereby obtaining decorrelated conformations of the RNA2 molecule at both salt concentrations (that is, 0.15 M and 0.5 M). As for this stage, MD simulations of $2\times 10^{8}$ steps were carried out at the constant temperature of $T = 310$ K, employing a timestep $\delta t = 3\times 10^{-3}\; \tau \simeq 9$ fs -- with $\tau$ defining the internal simulation time unit. The correlation time and timestep frequency of the Bussi thermostat were set to $1000$ and $53$ respectively -- as suggested by the software developers.

In addition, hydrogen-bonding interactions were switched off, to try and avoid that the RNA2 conformations were significantly biased by the relaxation pathway in the subsequent MD stages. We thus tracked the gyration radius ($R_g$) along the trajectories and extracted three, non-correlated RNA conformations per salt concentration, about the minimum (I), average (II), and maximum (III) values of $R_g(t)$ in the late plateau of the Bussi dynamics.

Secondly, we performed a Langevin MD simulation of each (six) selected RNA2 conformation ($T = 310$ K, $1\times10^{10}$ steps), hereby referred to as \textit{freely-folding} MD: Values of the diffusion coefficient and timestep frequency associated with the Langevin thermostat were set to $2.5$ and $103$ respectively, while the MD timestep $\delta t = 3\times 10^{-3}\; \tau$ was kept. At this stage, hydrogen-bonding and sequence-specific interactions were switched back on, thereby letting the systems fold into secondary and tertiary structural motifs.

\subsubsection*{Squeezing of the RNA2 molecule by a spherical confinement} \label{section:RNA2Packing_protocol}
To mimic a CCMV environment, the RNA2 molecule was first confined within a spherical volume that is compliant with the size of the inner cavity of the CCMV capsid. To this aim, we employed a radially-symmetric, time-dependent harmonic potential, acting effectively upon the nucleotides that lie about the enclosing surface, and defined as:
\begin{equation}\label{eq:HarmonicSpherePotential}
U_{hs}(r,t) = 
    \begin{cases} 
        0 & \text{if } r \leq R_{start} - \nu t , \\
        \frac{1}{2} \omega \left[  r - \left( R_{start} - \nu t \right) \right]^2 & \text{otherwise}
    \end{cases}
\end{equation}
where $r$ is the radial coordinate within a reference frame that is centered at the center of mass of the model capsid, $\omega = 1 \; k_B T \lambda^{-2}_{ox}$ defines the stiffness of the enclosing sphere (with $T = 310 K$ and the ``\textit{ox}'' subscript denoting the internal units employed by oxRNA2 -- refer to the Supplementary Material of \cite{oxrna2014} for details), and $\nu$ and $R_{start}$ are the encapsidation rate and starting radius associated with the squeezing regime respectively. All MD replicas were set up to achieve a final encapsidation radius between 10 and 12 nm.

MD simulations were performed in the NVT ensemble at $310$ K \textit{via} a Langevin thermostat: The diffusion coefficient and timestep frequency of the thermostat, and the simulation timestep of the freely-folding setup were kept. For the sake of the numerical efficiency, this squeezing protocol has been applied to those RNA conformations showing the lowest value of the gyration radius during the freely-folding stage within each (I, II, III) MD replica per salt concentration.

\begin{figure*}[ht]
    \centering
    \includegraphics[scale=0.25]{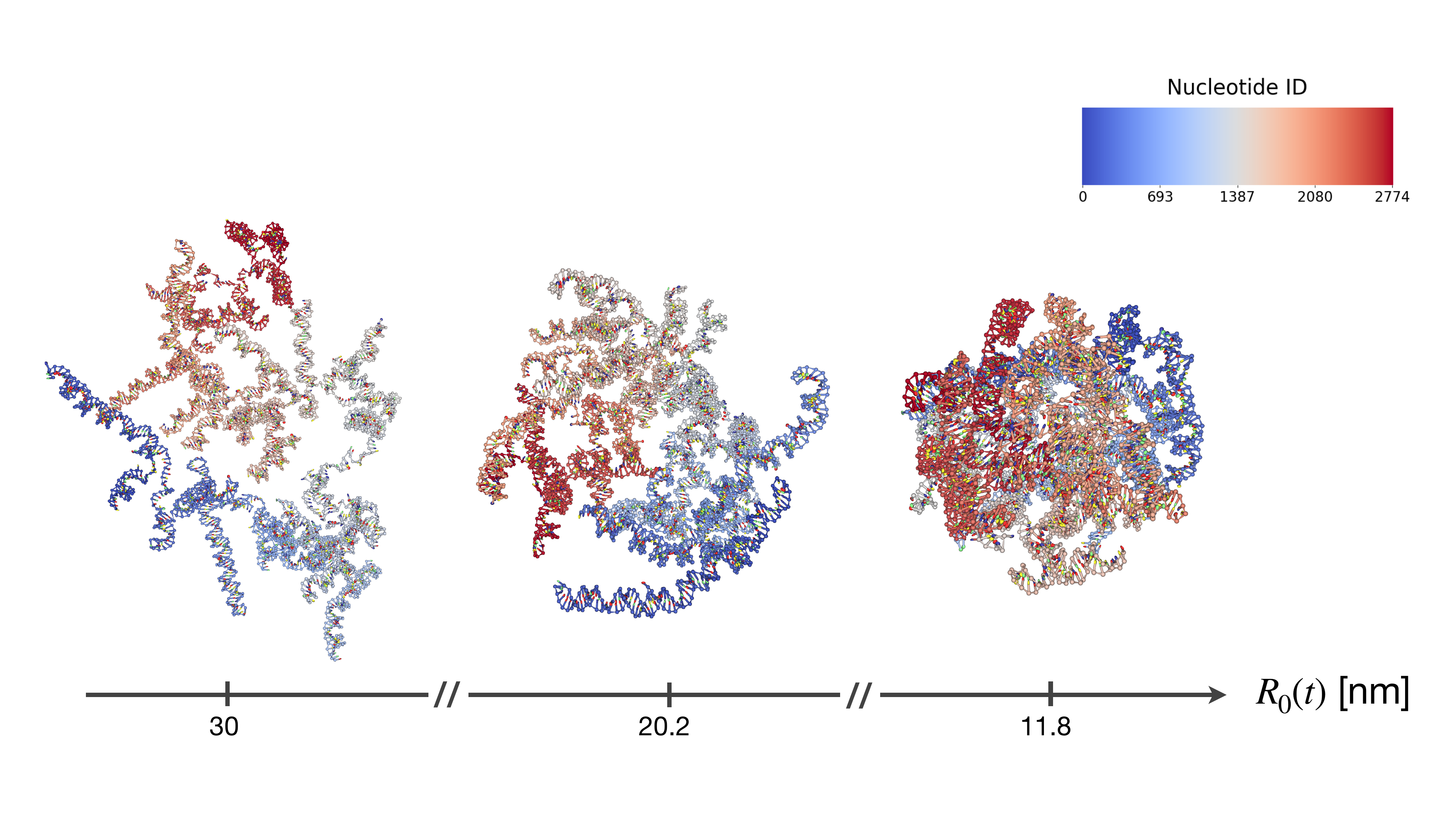}
    \caption{Snapshots of an example \textit{squeezing} stage for an RNA2 molecule at 0.5 M - values of the instantaneous radius of the enclosing sphere are reported below. The color scale is associated with the nucleotide sequence.}
    \label{fig:RNA2Packing_protocol}
\end{figure*}

For the enclosing potential to be the least disruptive of the molecular structure of the RNA2, we set a value of $\nu = 5.6\times 10^{-8}$ $\lambda_{ox} (\text{MD steps})^{-1}$, while making sure that the internal pressure of the system did not diverge throughout the simulations (see \textbf{Figure S10} of the Supplementary Material). As for the 0.5 M scenario, a satisfactory value of the encapsidation radius was achieved, exerting no excess external force on the RNA structure. Yet, to tolerate the squeezing stage, the 0.15 M RNA structures required a slightly milder procedure, i.e., with $\nu$ reset to $1\times 10^{-8} \; \lambda_{ox} (\text{MD steps})^{-1}$ about halfway of the MD trajectory. \textbf{Figure \ref{fig:RNA2Packing_protocol}} exemplifies the squeezing stage associated with RNA2 at 0.5 M.

\subsubsection{Dynamics of the encapsidated RNA2 within a mean-field external potential}\label{section:RNA2Packed_protocol}
The sole squeezing procedure yields an ensemble of confined, strained RNA2 conformations, which shall subsequently adjust to the electrostatic field of the CCMV capsid. To this concern, we adopted two mean-field approaches to mimic the inner environment of the CCMV cavity.

A first approach, denoted $U_{yuk+WCA}(r)$, is based on the analytical formulation reported by \v{S}iber and Podgornik~\cite{vsiber2007role} and subsequently adopted in other theoretical studies~\cite{vsiber2012energies}: The Poisson-Boltzmann equation is solved for the electrostatic potential associated with an empty icosahedral capsid, depicted as a spherical thin shell, at a fixed monovalent salt concentration. By employing the Debye-H{\"u}ckel approximation, an analytical solution is derived that applies to both the internal and the external environment of the viral capsid. 

\begin{figure*}[ht]
    \centering
    \includegraphics[width=0.6\linewidth]{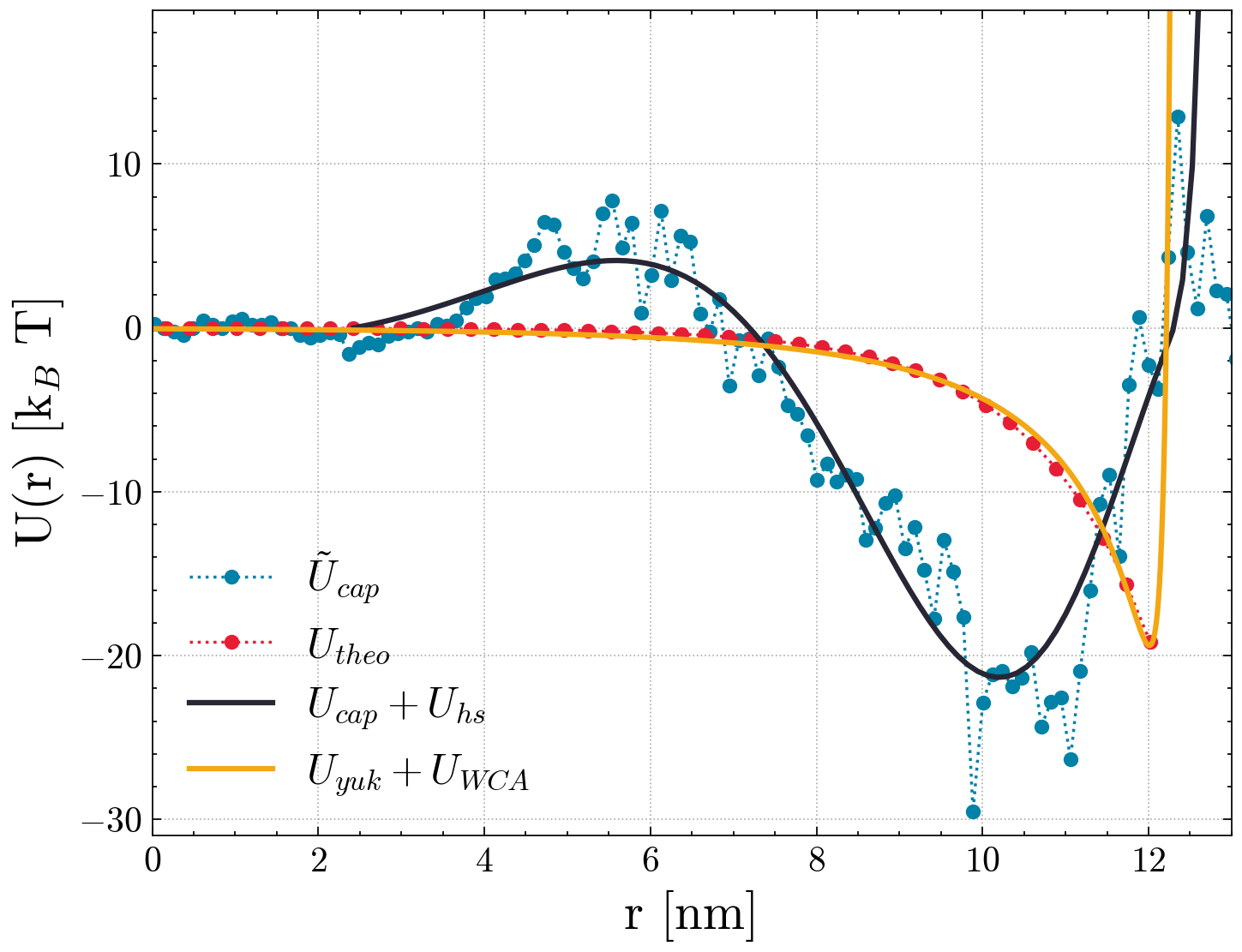}
    \caption{{Radial potential energy profiles of the CCMV capsid and the ionic shells thereof, which have been adopted for the dynamics of the encapsidated RNA2 (depicted as solid lines). Data points are associated with the analytical (red dots) and the structure-based approach (blue dots) respectively.}}
    \label{fig:CCMV_Potential}
\end{figure*}

Here, we fitted this solution \textit{via} a combination of a Yukawa attractive potential and a short-range WCA repulsive wall (see \textbf{Figure \ref{fig:CCMV_Potential}}), as:
\begin{multline}\label{eq:RudyPotential}
U_{yuk+WCA}(r) = U_{yuk}(r) + U_{WCA}(r) = \frac{\alpha \exp\left[ - (R_\delta-r)/\lambda\right]}{(R_\delta-r)}\\ + 4\epsilon\left[ \left( \frac{\sigma}{(R_\delta-r)}\right)^{12} - \left( \frac{\sigma}{(R_\delta-r)}\right)^6 \right] + \epsilon,
\end{multline}
with $R_\delta$ (a slightly higher value of) the radius of the CCMV cavity; $\alpha$ and $\lambda$ the amplitude and Debye length of the Yukawa potential; $\epsilon$ and $\sigma$ the amplitude and radius of the standard WCA potential. Parameters of the fitting procedure were implemented within the oxDNA simulation code and are reported in \textbf{Section IV.C }of the Supplementary Material.

\begin{figure*}[ht]
    \centering
    \includegraphics[width=0.8\linewidth]{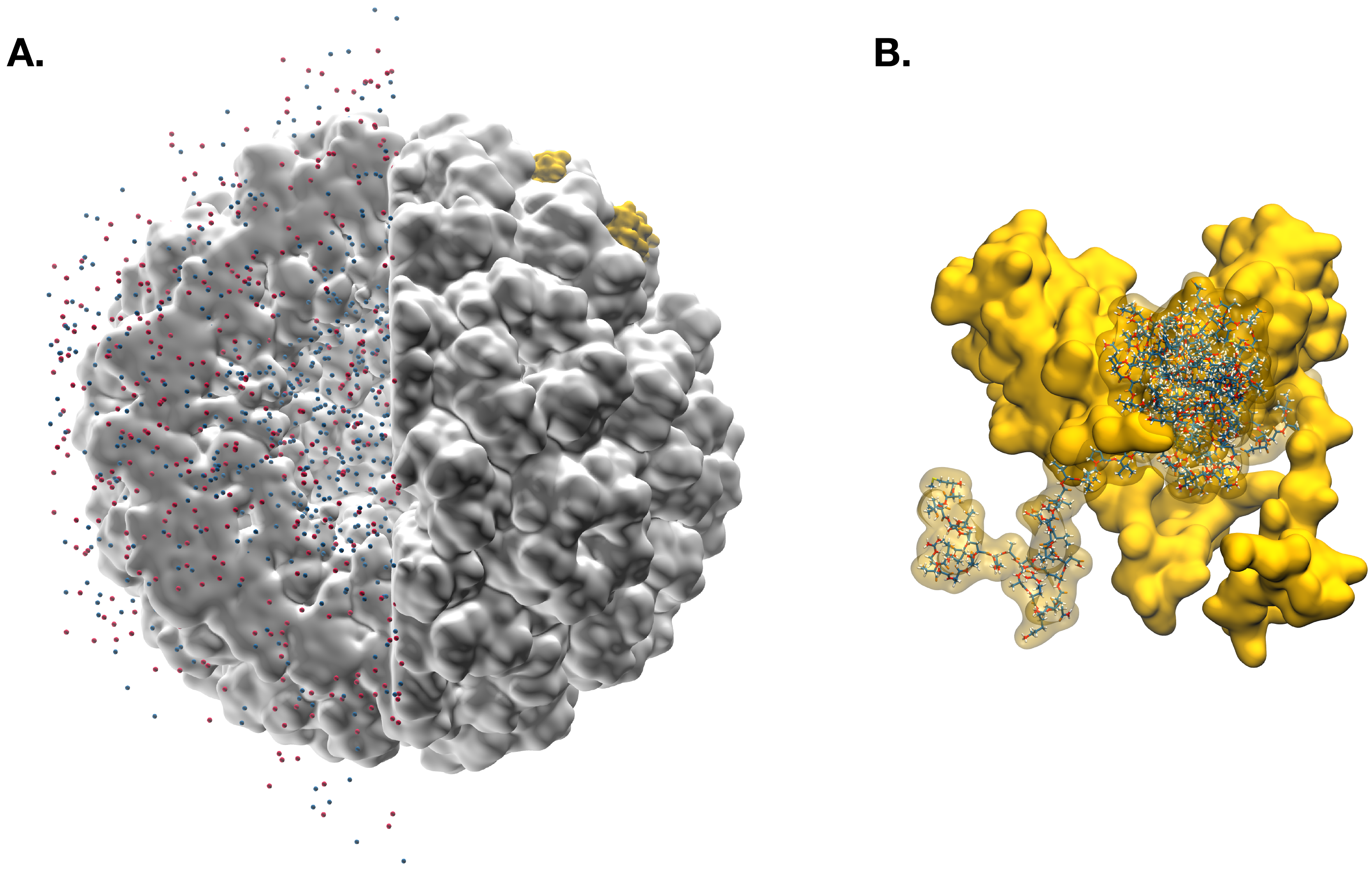}
    \caption{(\textbf{A}) Snapshot of the CCMV capsid and the thermalized shells of sodium (red) and chloride (blue) ions thereof. (\textbf{B}) Surface depiction of the trimeric symmetrical subunit of CCMV, along with the amino-terminal tails decorating the inner capsid walls.}
    \label{fig:CCMV_Structure}
\end{figure*}

The second approach, denoted $U_{cap}(r)$, was based on a multiscale protocol developed specifically for this work, whereby we employed the full structure of the CCMV capsid at the atomistic level of resolution to extract a mean-field, radial potential profile (details are discussed in \textbf{Section IV.B} of the Supplementary Material). Briefly: 
\begin{itemize}
\item upon the PDB template, the amino-terminal tails of the trimeric CCMV capsomer (i.e., the elementary symmetrical unit of the icosahedral capsid shell) were built \textit{via} the Modeller plugin \cite{fiser2003modeller} of the Chimera software \cite{pettersen2004ucsf}, and equilibrated both in vacuum and in solution (see \textbf{Figure \ref{fig:CCMV_Structure}\textbf{B}});
\item the newly-achieved trimer subunit was replicated according to the sixty-fold symmetry of the icosahedral shell, yielding an all-atom structure of the CCMV capsid;
\item the latter was subject to diverse rounds of energy minimization - both in vacuum and in solution, hence the solvent medium was thermalized \textit{via} a multi-step MD protocol. \textbf{Figure \ref{fig:CCMV_Structure}} shows a snapshot of the final conformation of the system.
\end{itemize}
We thus defined a radial potential energy profile upon the partial charge distribution of both the viral capsid and the thermalized ionic shells. To this purpose, we iteratively employed Gauss' theorem, relying on the approximation that the system was spherically-symmetric about the center of mass of the capsid. The charge distribution has been averaged over the solid angle and discretized along the radial coordinate, thereby achieving the values of $\Tilde{U}_{cap}[r]$ shown in \textbf{Figure~\ref{fig:CCMV_Potential}}. 

Lastly, we fitted $\Tilde{U}_{cap}[r]$ against an infinitely differentiable analytic function $U_{cap}(r)$ of the form:
\begin{equation}\label{eq:CCMVPotential}
    U_{cap}(r) = e^{A(R_0 - r)} \cdot \left( \sum^{N}_{n=1} c_n \cdot (R_0 - r)^n \right)
\end{equation}
with $N = 10$ and $R_0$ the radius of the inner CCMV cavity (likewise, parameters of the fit are reported in \textbf{Section IV.C} of the Supplementary Material). Additionally, we included a further harmonic potential about the capsid walls, as in \textbf{Equation~\ref{eq:HarmonicSpherePotential}} but with no time-dependence and $R_{start} = R_0$: this term mimics an excluded volume interaction, hence preventing the nucleotides from overcoming the fictitious capsid shell. Results of the fitting procedure are shown in \textbf{Figure \ref{fig:CCMV_Potential}}.

\subsection{Analysis protocol}
Analyses have focused on diverse structural features of the RNA2 configurational ensembles. Values of the hydrogen-bonding interactions, potential energy, and internal pressure of the RNA2 structure were calculated with the oxDNA software tools. Contact maps (KMs) were built on a frame-wise basis as Boolean matrices of the base pairings between nucleotides, whereby we tracked the evolution of the configurational ensembles of RNA2 and identified stable contacts, i.e., hydrogen-bonding interactions that are conserved over 50\% of the (total, individual) MD trajectories.

A distance criterion based on the Hamming distance between the the upper-triangular portion of two contact maps (i.e. counting the number of different entries) was adopted as metric of a hierarchical clustering - together with an average linkage criterion. Lastly, we relied on in-house scripts for all calculations and figures, based on oxDNA analysis tools \cite{poppleton2020design} and (scientific) Python libraries~\cite{harris2020array, virtanen2020scipy, gowers2019mdanalysis, scherer2015pyemma, shimoyama2022pycirclize, Hunter2007matplotlib, reback2020pandas}.

\section{Results and Discussions}

\subsection{Freely-folding dynamics of RNA2 in solution}\label{free-fold}
\begin{figure*}
    \centering
    \includegraphics[width=0.49\linewidth]{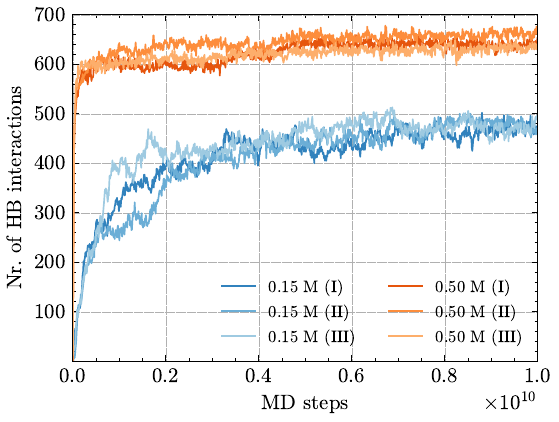}
    \includegraphics[width=0.49\linewidth]{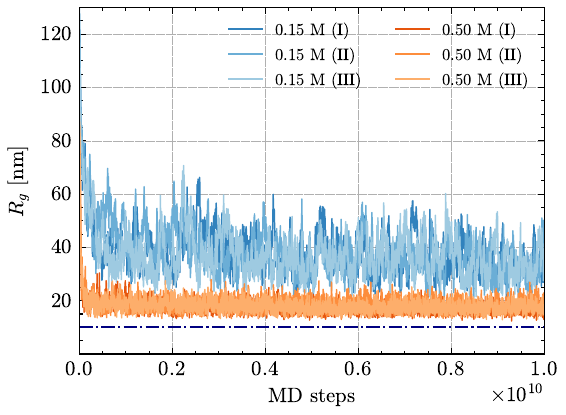}
    \caption{(\textbf{Left}) The total amount of hydrogen-bonding interactions and (\textbf{Right}) gyration radius of the RNA2 molecules, from the freely-folding MD at 0.15 M and 0.5 M: A less effective screening of the electrostatic repulsion accounts for a lower degree of nucleotide pairings and conformational compactness observed at 0.15 M. We note that all MD replicas per salt concentration converge to a consistent (HBs, $R_g$) plateau, regardless of the starting structure. The dot-dashed line displays the estimate value of the gyration radius of RNA2 reported in Ref.~\cite{marichal2021relationships}. }    \label{fig:NumberOfHB_RNARelaxation_AND_RadiusOfGyration_RNARelaxation}
\end{figure*}

Firstly, we characterized the behavior of a freely-folding RNA2 molecule in solution, at diverse monovalent salt concentrations -- i.e., 0.15 M and 0.5 M. As described in the Materials and Methods section, the RNA2 molecule has been subject to an equilibration protocol that allowed us to achieve an ensemble of non-correlated RNA configurations: This way, we would setup the subsequent freely-folding MD out of a diversified pool of starting structures, thus avoiding that the RNA2 molecules be stuck into locally-folded states. As hydrogen-bonding interactions are switched back on, the pairing of nucleotides drives a broad stabilization of the internal energy, together with a decrease in the gyration radius of the RNA molecules (as shown in \textbf{Figure \ref{fig:NumberOfHB_RNARelaxation_AND_RadiusOfGyration_RNARelaxation}}).
\newline
Overall, the $0.15$ and $0.5$ M scenarios display diverse conformational dynamics -- the latter showing a higher amount of hydrogen-bonding interactions and degree of compactness. This is in line with the Debye-H{\"u}ckel interactions screening the electrostatic repulsion less effectively at a lower salt concentration. We observe a discrepancy between the values of the gyration radii of RNA2 obtained here and the estimates of $R_{g} \sim 10$ nm reported by Marichal and co-workers from SAXS measurements in solution~\cite{marichal2021relationships}. This outcome might be accounted for by the different state point adopted in the experimental setup, with $T_{exp} = 293$ K and a nominal salt concentration about $0.1$ M. Yet, it is significant that all MD replicas per salt concentration converge to a consistent value of contacts (about 475 hydrogen bonds at $0.15$ M versus $650$ hydrogen bonds at 0.5 M) and gyration radius (about $35$ nm at $0.15$ M versus $17.5$ nm at $0.5$ M), regardless of the starting structure. 

Moreover, we note that the quality and kinetics of the folding process in the two scenarios are strongly salt-dependent. In fact, the trends observed at $0.5$ M highlight a steep rise in the amount of hydrogen-bonding interactions and a subsequent decrease in both the internal energy and the gyration radius of the RNA molecules, consistently leading to deep and narrow free energy basins (see \textbf{Figure S7} of the Supplementary Material). Notably, this is associated with a high variability of stable contacts, which are diversely distributed among MD replicas, as shown by the chord diagrams in both \textbf{Figure~\ref{fig:chords_all}} and \textbf{Figure S8} and \textbf{S9} of the Supplementary Material. We thus observe that the free RNA molecule at high ionic strength achieves a well-defined thermodynamic state, characterised by narrowly-distributed values of global observables, such as energy and gyration radius, which nonetheless emerge from a pool of rather diverse conformations at the microscopic, molecular level.

\begin{figure*}[ht]
    \centering
    \includegraphics[width=0.49\textwidth]{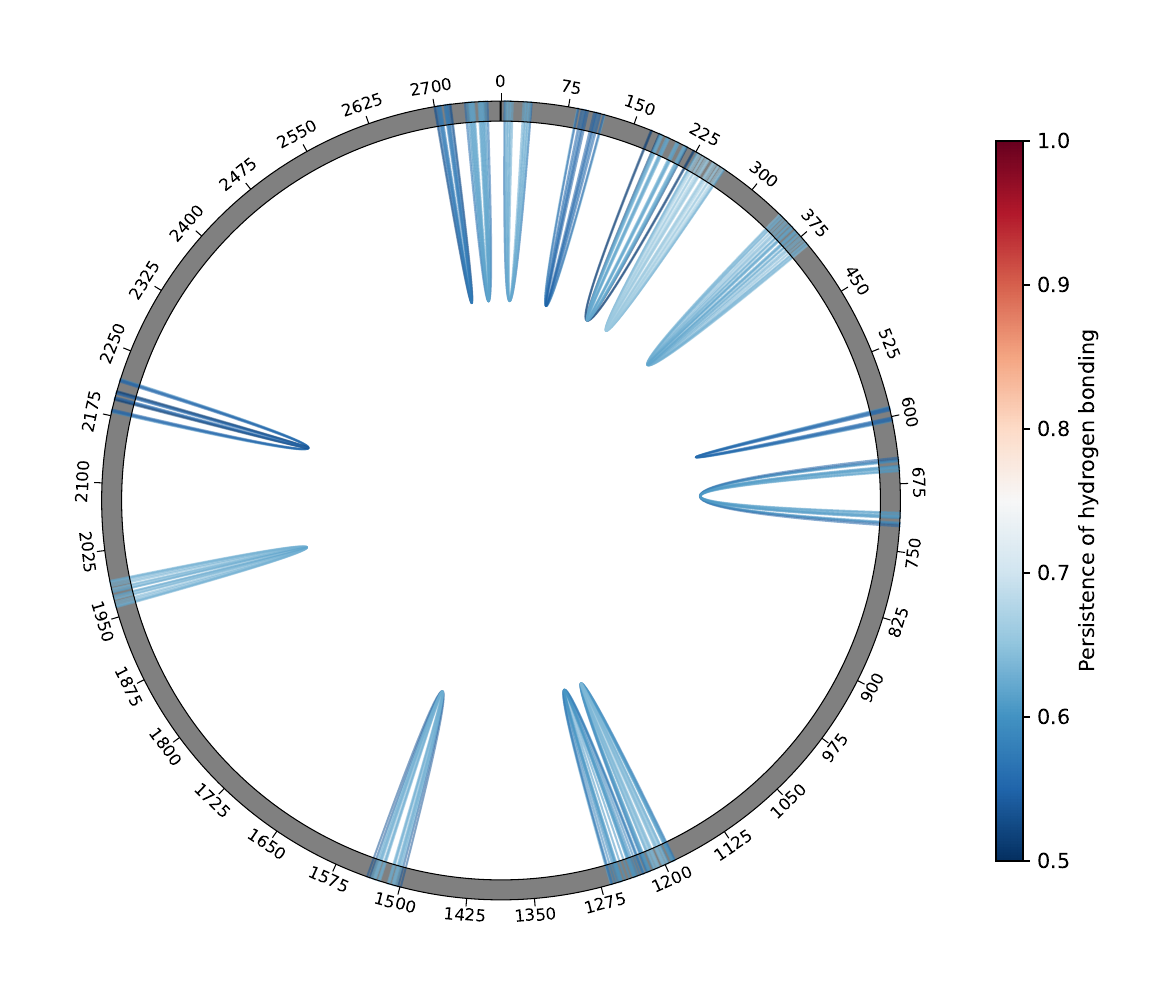}
    \includegraphics[width=0.49\textwidth]{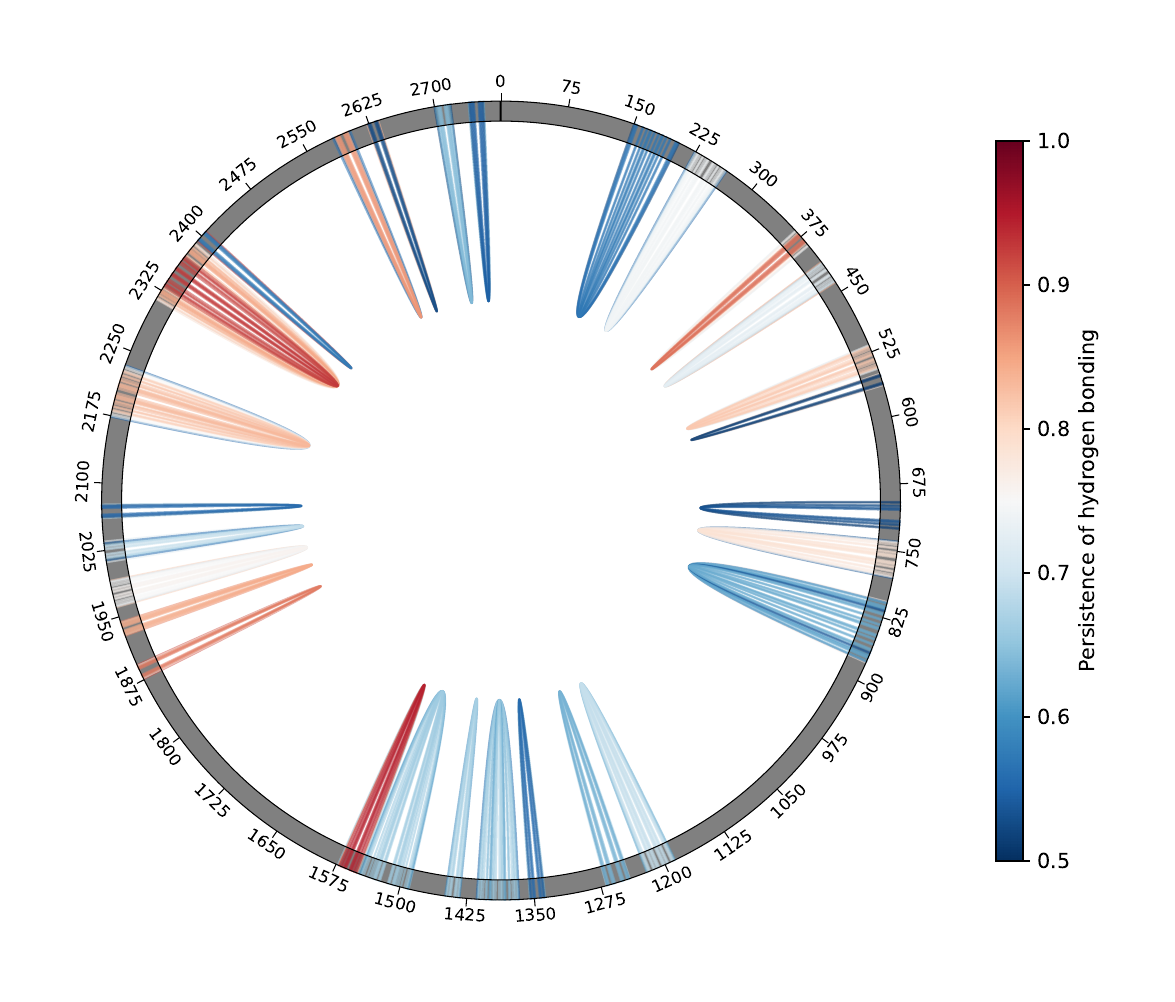}
    \caption{Chord diagram depiction of the stable hydrogen-bonding contacts (i.e., conserved in over $50\%$ of the total trajectory frames per salt concentration), from the equilibrated fraction of the freely-folding MD at (\textbf{left}) 0.5 M and (\textbf{right}) 0.15 M.}
    \label{fig:chords_all}
\end{figure*}

Conversely, at $0.15$ M a slower, steadier increase in the amount of hydrogen-bonding interactions accounts for broader and highly-degenerate free energy landscapes, whereby the RNA molecules have access to a range of states lying within $\Delta$R$_{g}$ = $10-15$ nm and $\Delta$U = $0.3-0.5$ k$_{B}$T. However, a non-negligible fraction of the stable contacts scored at $0.15$ M are shared between all MD replicas, as if all folding pathways consistently converged to an array of locally-folded secondary structures (see \textbf{Figure~\ref{fig:chords_all}} and \textbf{Figure S8} and \textbf{S9} of the Supplementary Material).

At the level of single MD replicas, hydrogen-bonding interactions between nucleotides span a wider range at $0.5$ M than at $0.15$ M, with the distribution of ladder distances shifting towards higher values in the former case. Nevertheless, we note that at neither salt concentration are established stable long-range interactions nor pseudoknots, which have been observed amongst functional RNA systems (such as rRNAs and tRNAs \cite{pleij1985new}) and whose occurrence is expected in viral RNA fragments \cite{brierley2007viral, yan2022length}.

These observations are corroborated by the analysis of the contact maps between nucleotides, which display a dynamical evolution in the structural motifs of the RNA2 molecules underneath a seemingly static equilibrium: In fact, at both salt concentrations, the systems progresses \textit{via} a steady re-distribution of the hydrogen-bonding interactions, as shown by both \textbf{Figure~\ref{fig:kM_distance_distribution}} and the distance matrices in \textbf{Section II.B} of the Supplementary Material.
\begin{figure*}[ht]
    \centering
    \includegraphics[width=0.32\linewidth]{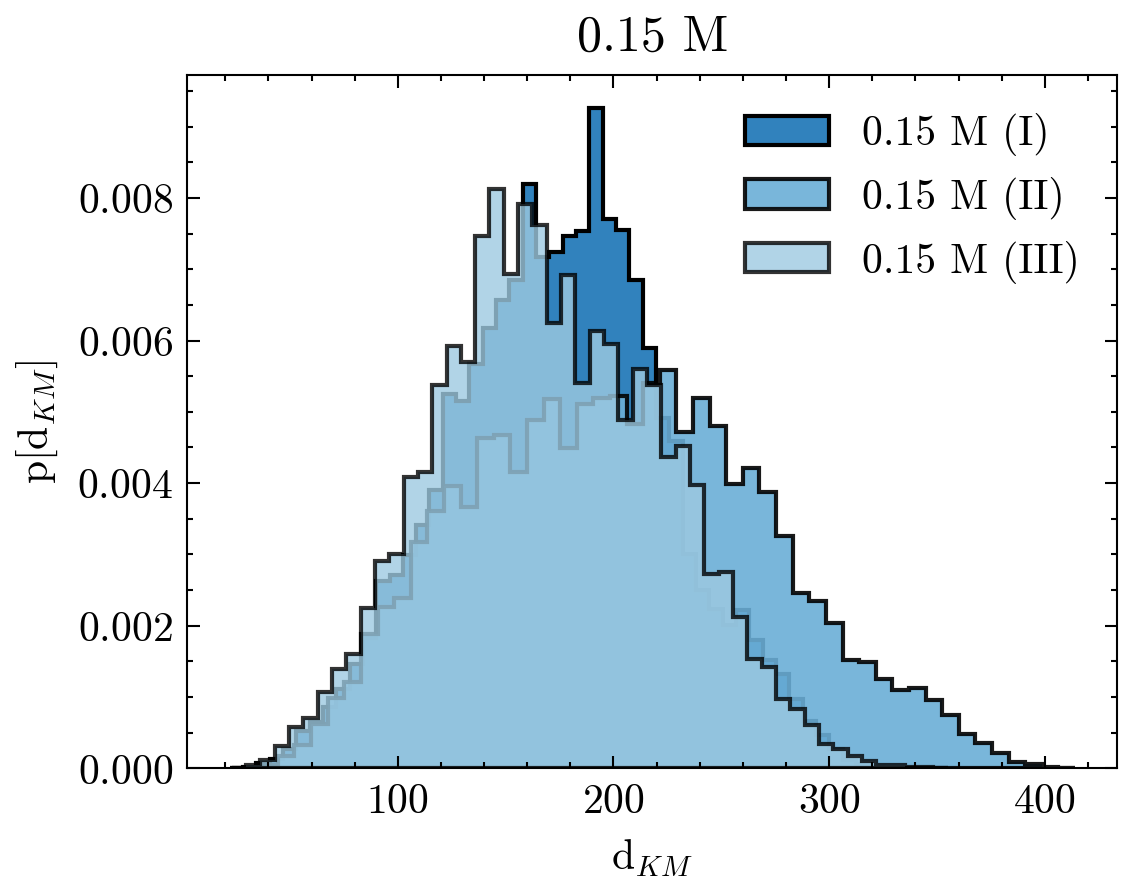}\hfill
    \includegraphics[width=0.32\linewidth]{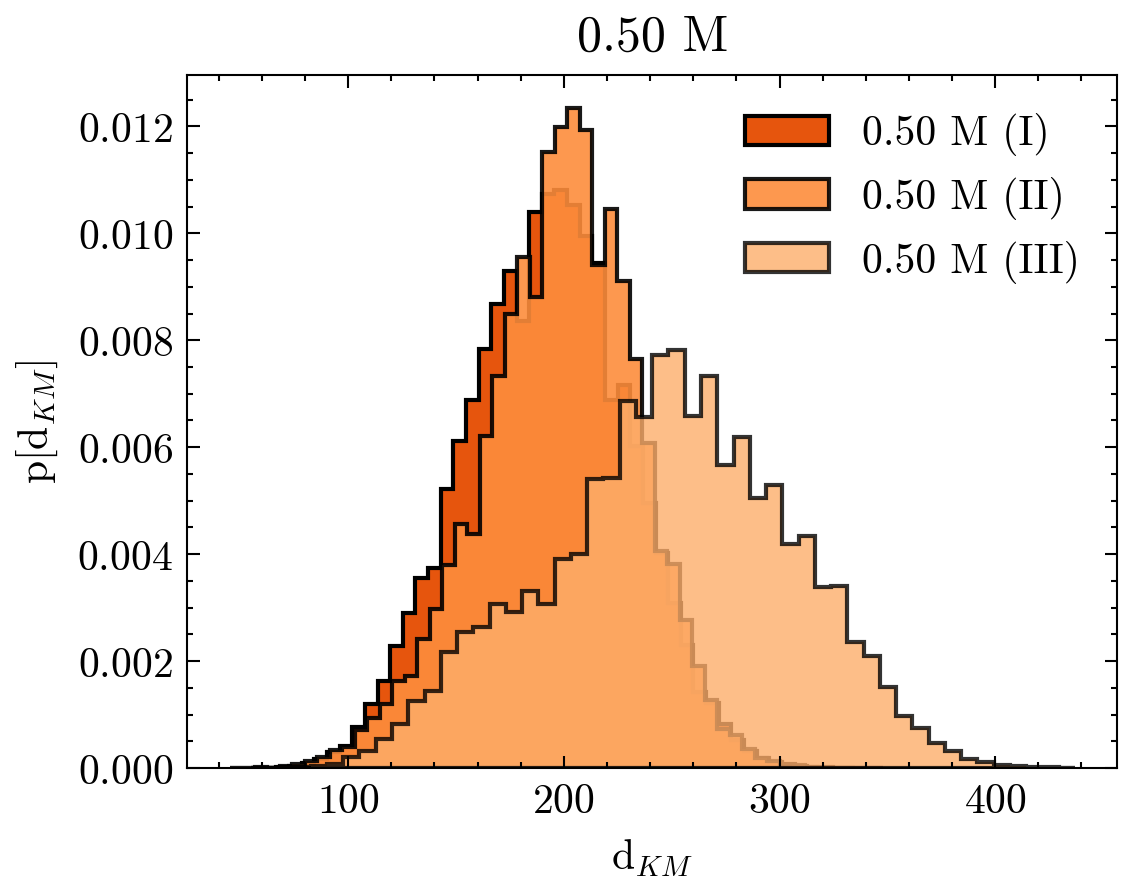}\hfill
    \includegraphics[width=0.32\linewidth]{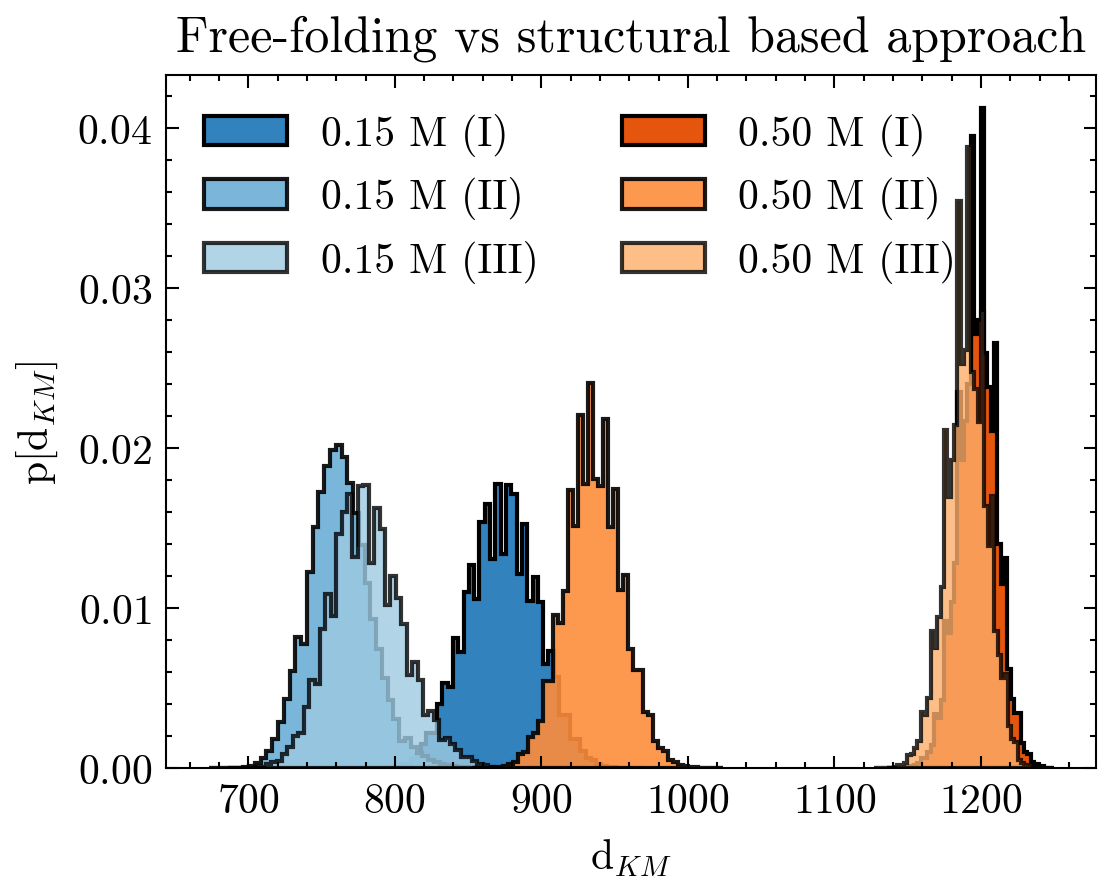}
    \caption{Normalized histograms of the distances between contact maps (defined in the ``Analysis protocol'' section), the latter being collected on a frame-wise basis for each replica of the freely-folding MD at (\textbf{left}) 0.15 M and (\textbf{center}) 0.5 M. \textbf{(Right)} Normalized histograms of the all-vs-all, cross-distances between the contact maps from the freely-folding MD and the contact maps from the dynamics of RNA2 within the CCMV-derived external potential: The ``shuffling'' of the secondary/tertiary structures enforced by the encapsidation procedure is highlighted.}
    \label{fig:kM_distance_distribution}
\end{figure*}
Moreover, these matrices, together with the hierarchical clustering of the contact maps (see the dendrograms in \textbf{Figure S5} and \textbf{S6} of the Supplementary Material), recapitulate our earlier observations on the non-trivial salt dependency of the conformational landscapes of RNA2: In fact, despite both scenarios exploring diverse structural ensembles (with no apparent overlap between secondary/tertiary motifs amongst MD replicas), the distances between contact maps at 0.5 M are about twice as high as they are at 0.15 M.

\subsection{Dynamics of RNA2 within a capsid-like electrostatic environment}\label{sec:RNA2PackedSection}
As detailed in the Materials and Methods section, the encapsidation of the RNA2 molecules involves two stages: i) the radial confinement of RNA2 (squeezing), to achieve molecular structures that are compliant with the size of the CCMV cavity -- associated with an inner radius $R_0 = 12$ nm, and ii) its adjustment to the external field of the CCMV capsid.

The external harmonic potential of the squeezing stage was fixed as a convenient trade-off between a quasi-static scenario -- that is, the least disruptive of the molecular topology -- and the numerical efficiency of the calculation (refer to \textbf{Section~\ref{section:RNA2Packing_protocol}} of the Materials and Methods).
Yet, we observed that the spatial proximity enforced by the squeezing procedure drives a significant rise in the amount of hydrogen-bonding interactions: This effect is steadier in the 0.5 M scenario and abrupt at 0.15 M, yielding a 60\% contact increase in the latter case (as shown in \textbf{Figure S11} of the Supplementary Material). 

Therefore, two mean-field approaches were employed to depict the electrostatic environment of the inner CCMV cavity, denoted $U_{yuk}(r)$ and $U_{cap}(r)$ (\textbf{Equations~\ref{eq:RudyPotential} and~\ref{eq:CCMVPotential}} in Section~\ref{section:RNA2Packed_protocol} of the Materials and Methods section, respectively). Overall, the outcome from the MD replicas employing either approach differs significantly. 
\begin{figure*}[ht]
    \centering
    \includegraphics[width=0.95\linewidth]{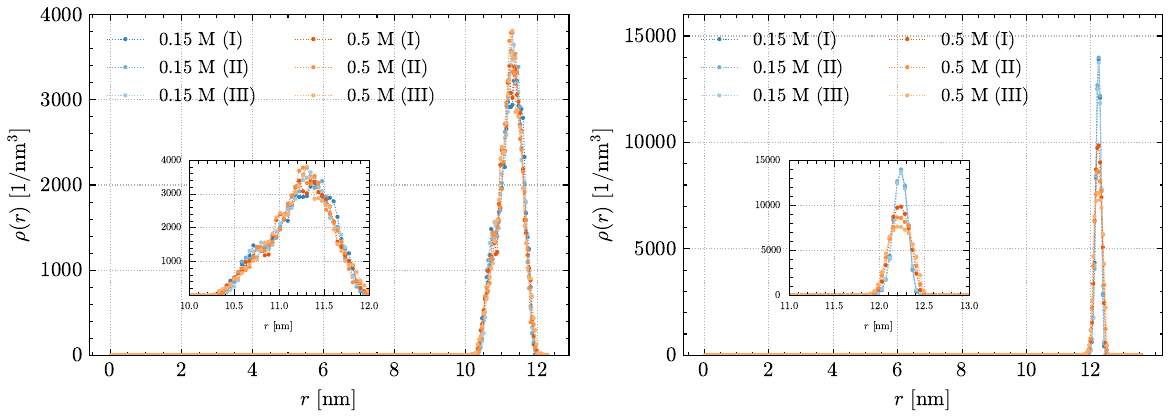}\hfill
    \caption{Profiles of the radial nucleotide density for the MD trajectories of RNA2 within a capsid-like electrostatic environment, derived \textit{via} mean-field approaches relying on either (\textbf{left}) atomistic data (structure-based approach), or (\textbf{right}) theoretical calculations (analytical approach) - details in the main text.}
    \label{fig:DensityOfNucleotides_RNAPacked}
\end{figure*}
In both cases, and at both salt concentrations, the RNA2 molecules adjust to the external field within about $1 \times 10^8$ MD steps, thereby (virtually) adhering to the internal walls of the CCMV capsid. Yet, the radial distribution of nucleotides (\textbf{Figure \ref{fig:DensityOfNucleotides_RNAPacked}}) is broader in the $U_{cap}(r)$ scenario, whereas $U_{yuk}(r)$ confines RNA2 to a narrow layer about the inner capsid walls. We note that the former is qualitatively in line with both the theoretical calculations reported by Marichal and co-workers~\cite{marichal2021relationships} and with the data of Zhang and co-workers~\cite{zhang2004electrostatic}, thereby highlighting the critical role of the N-terminal tails in the description of the electrostatic environment.

\begin{figure*}[ht]
    \centering
    \includegraphics[width=0.9\linewidth]{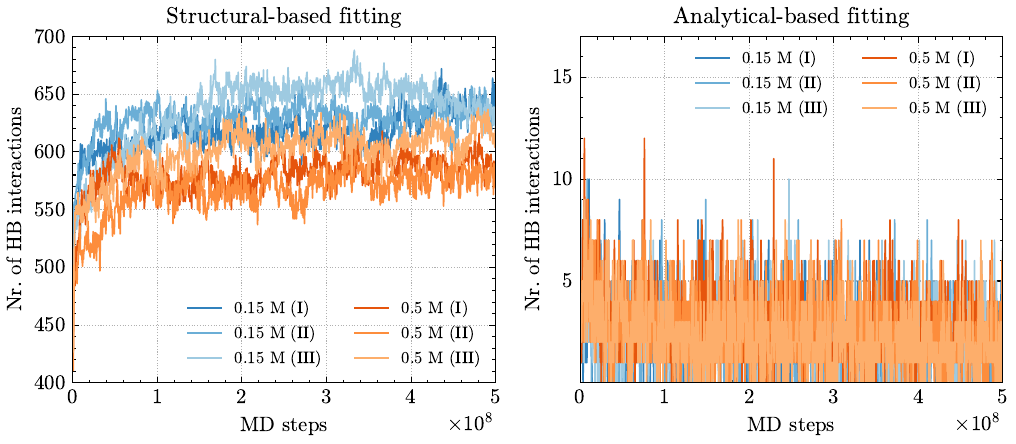}
    \caption{The total amount of hydrogen-bonding interactions from the MD trajectories of RNA2 within a capsid-like electrostatic environment, obtained by applying either (\textbf{left}) a structure-based approach ($U_{cap}(r)$) or (\textbf{Right}) an analytical approach ($U_{yuk}(r)$) (details in the text).}
    \label{fig:NumberOfHB_PACKED}
\end{figure*}

\begin{figure*}[ht]
    \centering
    \includegraphics[width=0.5\linewidth]{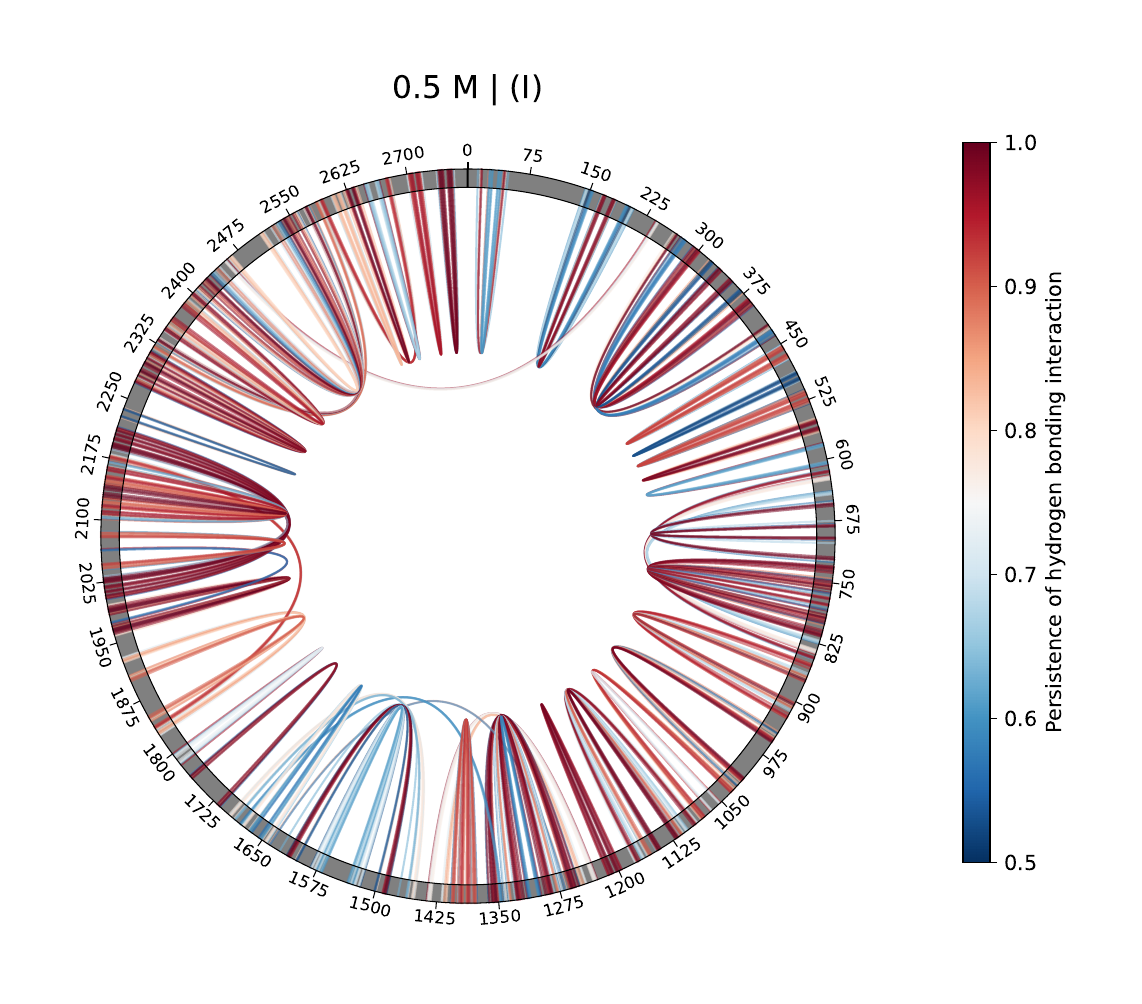}\hfill
    \includegraphics[width=0.5\linewidth]{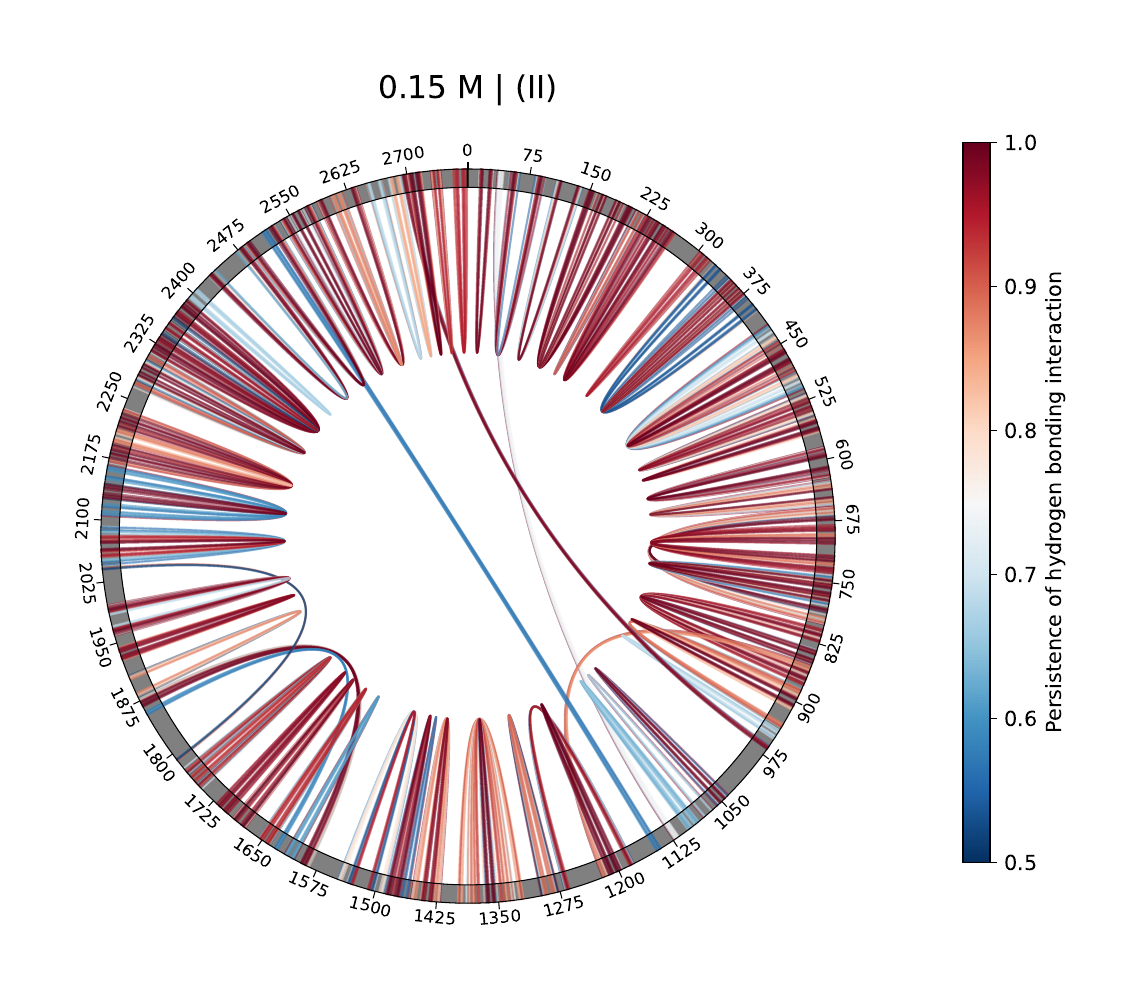}
    \caption{Chord diagram depiction of the stable hydrogen-bonding contacts, i.e., conserved in over $50\%$ of the trajectory frames, corresponding to (\textbf{left}) replica I at 0.5 M, and (\textbf{right}) replica II at 0.15 M, within the $U_{cap}(r)$ capsid-like, electrostatic environment (that is, obtained \textit{via} the structure-based approach -- details in the main text).}
    \label{fig:chords_PACKED}
\end{figure*}

In fact, the $U_{yuk}(r)$ external potential enforces a systematic disruption of nucleotide pairings and RNA secondary structures (see \textbf{Figure~\ref{fig:NumberOfHB_PACKED}}) -- which tallies, however, with the discussion in~\cite{erdemci2016effects}. \\
Not only is this behavior not recapitulated by applying $U_{cap}(r)$, but the amount of hydrogen-bonding interactions builds up in the latter case, as shown by \textbf{Figure~\ref{fig:NumberOfHB_PACKED}}. Moreover, within the external field of the $U_{cap}(r)$ CCMV capsid, the $0.15$ M scenarios slightly outrank the $0.5$ M counterparts over the total amount of nucleotide pairings, thereby suggesting that the CCMV environment might outcompete the electrostatic stabilization of the saline medium. In line with this observation, the internal energy difference between the $0.15$ M and $0.5$ M MD replicas drops to about $0.5\; k_{B}T$ units (see \textbf{Section IV.D} of the Supplementary Material).
The structural enhancement driven by $U_{cap}(r)$ yields a rich variety of contacts and long-range motifs, such as RNA pseudoknots, in both the 0.15 M and 0.5 M scenarios (as shown by \textbf{Figure~\ref{fig:chords_PACKED}}). As observed earlier (\textbf{Section~\ref{free-fold}}), several stable contacts are systematically found in all MD replicas at 0.15 M (see \textbf{Figure S19} of the Supplementary Material). Notably, a significant fraction of these contacts is identically found in the freely-folding trajectories: Thus, despite the squeezing procedure significantly scrambling the ensembles of conformations obtained in the freely-folding stage (as shown by the right panel in \textbf{Figure \ref{fig:kM_distance_distribution}}), the RNA2 is seemingly capable of conserving a few robust secondary motifs, in a strongly salt-dependent manner.

\section{Conclusions}
In this work, we have explored the folding dynamics and conformational ensemble of the RNA2 fragment of the cowpea chlorotic mottle virus, by means of CG molecular dynamics and mean-field approaches that explicitly take into account the electrostatic environment of the inner cavity of the capsid.

In the freely-folding context, all the diverse MD replicas converge to consistent values of the gyration radii and amount of hydrogen-bonding interactions. In fact, the RNA2 displays a dynamical equilibrium at both values of the saline concentrations explored, namely $0.15$ M and $0.5$ M, whereby the distribution of nucleotide pairings and secondary structures swiftly varies throughout the MD trajectories and between MD replicas. Notably, the RNA2 molecule seemingly relies on an array of short-ranged, stable contacts consistently found in all MD replicas at $0.15$ M and to a lesser extent at $0.5$ M, regardless of the starting conformation. The gyration radii obtained from the ensembles of MD configurations of the freely-folding scenario does not agree with the currently available experimental estimates \cite{marichal2021relationships}. This discrepancy might be ascribed to a number of factors, ranging from the significantly different temperatures of the \textit{in vitro} and \textit{in silico} setups, to the intrinsic limitations of the oxRNA model here employed, which in fact lacks an explicit treatment of the saline medium and a proper description of several non-canonical interactions.

Subsequently, a two-steps encapsidation protocol was enforced: first, we confined the RNA2 molecules into a spherical volume compliant with the inner cavity of CCMV; second, we allowed the RNA to adjust to the electrostatic field within the capsid. For the latter, two mean-field approaches were adopted, based on either the theoretical formalism reported by \v{S}iber and Podgornik~\cite{vsiber2007role} (analytical approach), or derived from atomistic structures of the inner walls and N-terminal tails of CCMV and the ionic distribution thereof (structure-based approach).

Despite these alternative formulations achieving somewhat analogous potential profiles, their impact on the behavior of the encapsidated RNA2 differs substantially. In fact, the analytical approach clears all traces of secondary structures and nucleotide pairings obtained in the freely-folding scenario. In contrast, the structure-based approach significantly changes the contact pattern of RNA2 while yielding a rich variety of medium- to long-range motifs and pseudoknots, which were not observed in a freely-folding stage. This is in line with Ref.~\cite{larman2017packaged}, where different secondary structures of packaged and free RNA have been experimentally observed in STMV. Yet, a significant fraction of the stable contacts observed at $0.15$ M is found in all MD replicas of both the free and the encapsidated setups, further highlighting a critical role of the saline medium to driving sequence-specific and kinetic effects. These results highlight that the variability of secondary structures is substantial in both regimes as expected~\cite{yoffe2008predicting,gopal2014viral,tubiana2015synonymous,vaupotivc2023viral}, although further validation is required.

Arguably, the description of the CCMV electrostatic field, which enforces a significant bias upon the conformational ensembles explored by RNA2, is critically sensitive to the charge distribution of the inner capsid walls and the ionic shells thereof~\cite{perlmutter2013viral,dong2020effect,bruinsma2021physics}.
In this respect, our results support the hypothesis that a proper description of the electrostatic contribution from the N-terminal tails of the capsid, even only at a mean-field level, is required to achieve stable RNA secondary structures \emph{in virio}.
This is in line with earlier CG numerical studies, which relied on the explicit inclusion of the charge carried by the tails to explain the fact that viral RNA is overcharged~\cite{perlmutter2013viral}.

We remark that, by employing effective approaches, we are limited in our capability to infer about the actual folding and packaging pathways of RNA2: In fact, both processes are influenced by a variety of co-factors \textit{in vivo} and in cultured cells experiments~\cite{comas2019packaging}, although CCMV was demonstrated to spontaneously yield supramolecular assemblies \textit{in vitro} that are indistinguishable from the native, infectious particle~\cite{fox1998comparison}. Moreover, the two-step, \textit{in vitro} assembly of the CCMV virions likely follows an \textit{en masse}-adsorption mechanism~\cite{elrad2010encapsulation, cadena2012self}, whereby a stoichiometric excess of capsid subunits is reversibly adsorbed onto the RNA chain, only to coalesce into a fully-mature virus-like particle upon lowering the pH. 
This notwithstanding, numerical techniques might be of aid in the characterization of viral RNAs: For instance, with no \textit{a priori} knowledge on the encapsidated RNA structures, Larsson and van der Spoel have inferred that the \textit{in silico} distribution of chloride ions well-recapitulates the electron density of the ssRNA in STMV and STNV~\cite{larsson2012screening}, thereby implicitly suggesting that a proper effective potential should be derived \textit{via} a self-consistent protocol, taking into account the very distribution of the RNA molecule. Similarly, this work demonstrates the capabilities of the numerical tools at our disposal to characterize the inherent structural variability of viral RNAs: The combination of molecular dynamics with a multi-scale modelling approach defines a robust setup to achieve a quantitative description of a capsid-like environment, taking into account the explicit contribution of the electrostatic forces at an apt level of resolution. As such, this technique might support the conventional approaches based on thermodynamic structure predictions, in the perspective of a proper mechanistic depiction of the assembly process.

\newpage

\medskip
\textbf{Supporting Information} \par 
Supporting Information is available from the Wiley Online Library or from the author.

\medskip
\textbf{Data Availability} \par 
The raw data produced and analyzed in this work are freely available on a Zenodo repository at the web address \href{https://doi.org/10.5281/zenodo.13221763}{doi.org/10.5281/zenodo.13221763}.

\medskip
\textbf{Funding} \par 
LP, MM, and RP acknowledge support from Fondazione CARITRO through the project COMMODORE (\#20260).

LP, MM, and RP acknowledge support from the Italian Ministry of Education, University and Research (MIUR) through the FARE grant for the project HAMMOCK (grant R18ZHWY3NC).

RP acknowledges funding from the European Research Council (ERC) under the European Union’s Horizon 2020 Research and Innovation Programme (Grant Agreement 758588).

LT and RP acknowledge support from ICSC – Centro Nazionale di Ricerca in HPC, Big Data and Quantum Computing, funded by the European Union under NextGenerationEU. Views and opinions expressed are however those of the author(s) only and do not necessarily reflect those of the European Union or The European Research Executive Agency. Neither the European Union nor the granting authority can be held responsible for them.

Funded by the European Union under NextGenerationEU, PRIN 2022 Prot. n. 2022Z3BBPE (SIGMA-1) and 2022RYP9YT (SCOPE).

This work was partially supported by the French National Research Agency (MERLIN ANR-22-CE45-0032).

\medskip

\textbf{Acknowledgements} \par 
We acknowledge the CINECA award under the ISCRA initiative (Project ID: HP10CGJ627) for the availability of high-performance computing resources and support.

GM, MM, and LP wish to thank Vittorio Chiavegato for providing the fitting procedures employed throughout the work, and Prof. Luca Monticelli for an extremely insightful conversation on the artificial effects emerging from the atomistic simulations of viral capsids. All authors wish to acknowledge the work and support of Prof. Lorenzo Rovigatti, who promptly provided the oxDNA code with a CUDA implementation of the Yukawa external force, performed benchmark, freely-folding simulation of the thermal annealing of RNA2, and engaged us in extremely enjoyable scientific conversations.

\medskip

\bibliography{main}

\end{document}